\documentclass[pre,onecolumn,preprintnumbers,floatfix,amssymb,amsmath]{revtex4}
\usepackage{graphicx}
\usepackage{natbib}
\usepackage{subfigure}
\usepackage{color}
\usepackage{multirow}
\usepackage{amssymb}

\begin{document}

\title{Drop pattern resulting from the breakup of a bidimensional grid of liquid filaments}
\author{Ingrith Cuellar, Pablo D. Ravazzoli, Javier A. Diez and Alejandro G. Gonz\'alez}
\affiliation{Instituto de F\'{\i}sica Arroyo Seco, Universidad Nacional del Centro de la Provincia de Buenos Aires, and CIFICEN-CONICET-CICPBA, Pinto 399, 7000, Tandil, Argentina}

\begin{abstract}
A rectangular grid formed by liquid filaments on a partially wetting substrate evolves in a series of breakups leading to arrays of drops with different shapes distributed in a rather regular bidimensional pattern. Our study is focused on the configuration produced when two long parallel filaments of silicone oil, which are placed upon a glass substrate previously coated with a fluorinated solution, are crossed perpendicularly by another pair of long parallel filaments. A remarkable feature of this kind of grids is that there are two qualitatively different types of drops. While one set is formed at the crossing points, the rest are consequence of the breakup of shorter filaments formed between the crossings. Here, we analyze the main geometric features of all types of drops, such as shape of the footprint and contact angle distribution along the drop periphery. The formation of a series of short filaments with similar geometric and physical properties allows us to have simultaneously quasi identical experiments to study the subsequent breakups. We develop a simple hydrodynamic model to predict the number of drops that results from a filament of given initial length and width. This model is able to yield the length intervals corresponding to a small number of drops and its predictions are successfully compared with the experimental data as well as with numerical simulations of the full Navier--Stokes equation that provide a detailed time evolution of the dewetting motion of the filament till the breakup into drops. Finally, the prediction for finite filaments is contrasted with the existing theories for infinite ones.
\end{abstract}

\maketitle

\section{Introduction}
\label{sec:intro}

The synthesis and assembly of structures at the nanoscale play a crucial role in many fields of technological and scientific interest~\cite{fleming_pt08}. One of the fields where both self and directed assemblies are relevant is the generation of nanoscale metallic particles which can be used as a basis for controlled growth of carbon nanofibers~\cite{henley_ass07} that are fundamental in numerous settings~\cite{fan_sc2008}. In a more general fashion, the formation of nanostructures of metallic materials play a significant role in fields that range from plasmonics to liquid crystal displays and solar cells~\cite{maier_nm03,sun_sc2000}. For example, the size and distribution of metallic particles affects the coupling of surface plasmons with incident electromagnetic energy. Thus, it is expected that the yield of solar cell devices can be largely increased by controlling this coupling~\cite{maier_b2003}. This particular case serves to illustrate the wide technological importance of being able to build uniformly distributed and controlled spaced metallic nanoparticles~\cite{baderi_rmp2006,min_nm2008}. 

One approach to produce desired nanoscale structures (metallic or not) with prescribed size and distribution is to resort to naturally occurring forces that drive the evolution of instabilities in the liquid phase~\cite{fowlkes_nl14} from an initially patterned nanostructure. Such an approach, if conveniently controlled, is significantly more efficient than lithographically depositing individual particles. For metal films, a recently developed technique consists in the fast liquefaction with pulsed lasers of an initial solid metallic film that has been previously shaped by electron beams. The metal film becomes unstable and breaks up into droplets while in the liquid phase, and they solidify later on remaining on the substrate as solid particles.  However, one difficulty in the study of the nanometric scale processes is that it is very difficult to have a detailed temporal description of the breakup process. The same type of phenomena is observed in macroscopic films and filaments deposited on substrates partially wet by a liquid. Although there are obvious differences between experiments at such different scales, there are also some common features that have been done fruitfully analyzed in previous works~\cite{hartnett_lang15}. On the other hand, the millimetric experiments can be studied with reasonable detail in time and may provide a good benchmark for the models proposed to describe the rupture process. Therefore, the understanding of this type of process for submillimetric scales can provide useful insights on the processes underlying the ruptures. The focus of this paper is to understand the patterns of drops self--assembled in grids of fluid filaments and the way they are produced from the hydrodynamic point of view. 

In our study, the hydrodynamic unstable evolution of liquid filaments deposited on a solid substrate gives rise to a characteristic geometrical disposition of droplets which is a consequence of both the initial conditions and the subsequent natural development of ruptures. This can be related to  experimental setups~\cite{fowlkes_nano11,hartnett_lang15} recently developed at the nanometric scale that considers a geometry consisting of thin  strips with thicknesses of tens of nanometers, widths of hundreds of nanometers, and lengths of tens of microns. When liquefied by laser pulses of typical duration of tens of nanoseconds (PLiD), these strips quickly retract into filaments that then break up into droplets. The droplets spacing is not uniform, but obeys a spatial distribution which is consistent with the prediction of a stability analysis of the Rayleigh–Plateau (R–-P) type. This mechanism of breakup can be explained surprisingly well by an analogy with the R–-P analysis of the breakup of a free standing fluid jet, modified by the presence of substrate~\cite{kd_pre09}. It has been proved~\cite{fowlkes_nano11} that by varying the width of the deposited metal strip with a sinusoidal perturbation of a well defined wavelength the dewetting process yields an array of uniformly spaced particles, as long as this imposed wavelength is unstable in the R–-P instability analysis. It has also been shown that perturbing with stable (short) wavelengths leads to distances between the particles that are nonuniform and not related to the imposed perturbation. 

In this paper, we focus on the description of the final drops pattern that results from a given rectangular grid of liquid filaments. The latter is characterized by the distance between the parallel filaments and its width, $w$. In Section~\ref{sec:exp} we describe the experimental setup to generate the grid, and discuss the main features of the liquid and substrate used. In a first stage, the grid rapidly breaks up around the nodes leading to a set of shorter filaments along the sides of the rectangles as well as drops at the crossings. In a later stage, the filaments start a retraction process from their ends~\cite{rava_pre17} and, finally, each one breaks up into a certain number of drops. Interestingly the drops formed at the crossings and those formed along the filaments have a different morphology. Thus, we devote Section~\ref{sec:drops} to study in detail the different geometrical features of each type of drops. In Section~\ref{sec:fil} we analyze how the spacing and distribution of the final droplets depend on the initial length of the filament, $L_i$. We study the relationship between the number of drops and $L_i$ by developing a simple hydrodynamic model, and compare its predictions with the experiments as well as with numerical simulations of the full Navier--Stokes equation (see Section~\ref{sec:num}). In Section~\ref{sec:conclu}, summarize the results and we elaborate on applications to nanoscale configurations.

\section{Experimental setup}
\label{sec:exp}

The experiments were carried out by producing liquid filaments with a silicone oil (polydimethylsiloxane, PDMS), which are placed on a substrate that they partially wet. The substrate is a microscope slide (glass) which is coated by immersion in a fluorinated solution (EGC-1700 of 3M) under controlled speed using a Chemat Dip Coater. This process ensures that the PDMS partially wets the substrate, since the solidified EGC--1700 coating lowers the surface energy of the glass. In order to have reproducible wetting properties, and to get rid of the solvent remaining in the coating, the coated substrates are left for a pair of days till the solvent is evaporated and the properties of the coating stabilized.
The detailed wettability properties of the PDMS on these substrates have been measured previously~\cite{rava_pof16,rava_pre17}. The wetting phenomenology is of the hysteretic type, and it is characterized by the advancing and receding (static) contact angles, $\theta_a$ and $\theta_r$, respectively. For the experiments in this work, we have $\theta_a=52^\circ$ and $\theta_r=44^\circ$.

Both the surface tension, $\gamma$, and density, $\rho$, of the PDMS are measured with a Kr\"uss\ K11 tensiometer, while its viscosity, $\mu $, is determined with a Haake VT550 rotating viscometer. The values obtained for these parameters are: $\gamma = 21.0$~dyn/cm, $\rho = 0.97$~g/cm$^{3}$, and $\mu = 90.7$~Poise at temperature $T=23^\circ $C.

\begin{figure}[htb]
    \centering
    \includegraphics[width=0.4\textwidth]{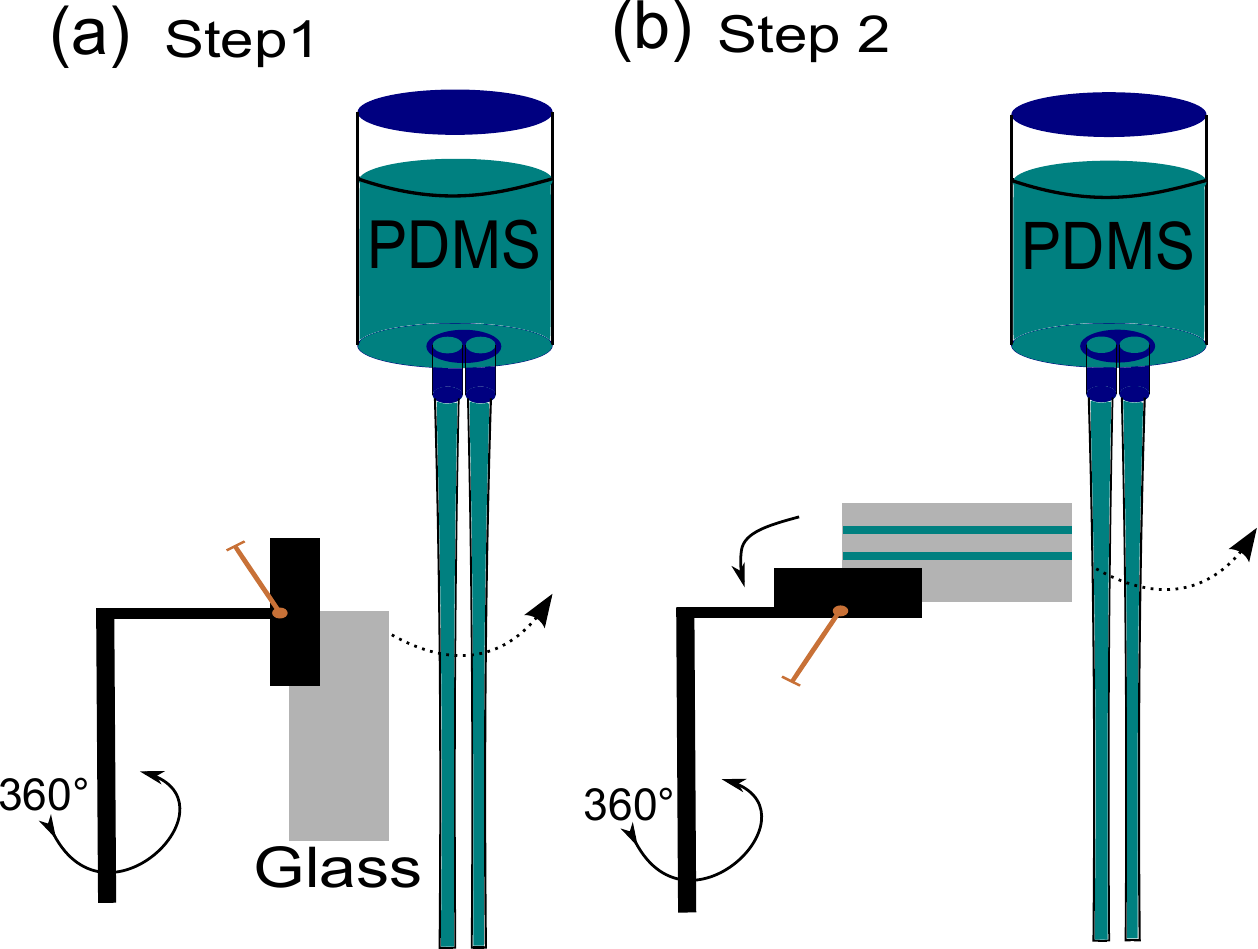}
    \caption{Experimental setup to produce and capture the pairs of filaments on the substrate in two steps.}
    \label{fig:setup}
\end{figure}

In order to study the formation of different types of drops, as well as the evolution of short filaments, we develop a particular configuration which consists of a pair of parallel long filaments that are crossed at right angle by another pair of filaments. We achieve this geometry by firstly capturing the filaments from two jets flowing down a vessel full with PDMS. This is done by rotating the substrate holder $360^\circ$ around a vertical axis (see Fig.~\ref{fig:setup}). Quickly afterwards, the holder is rotated $90^\circ$ around a horizontal axis, and spun again around the vertical axis to capture another pair of filaments.

\begin{figure}[htb]
    \centering
    \includegraphics[width=0.25\textwidth,angle=90]{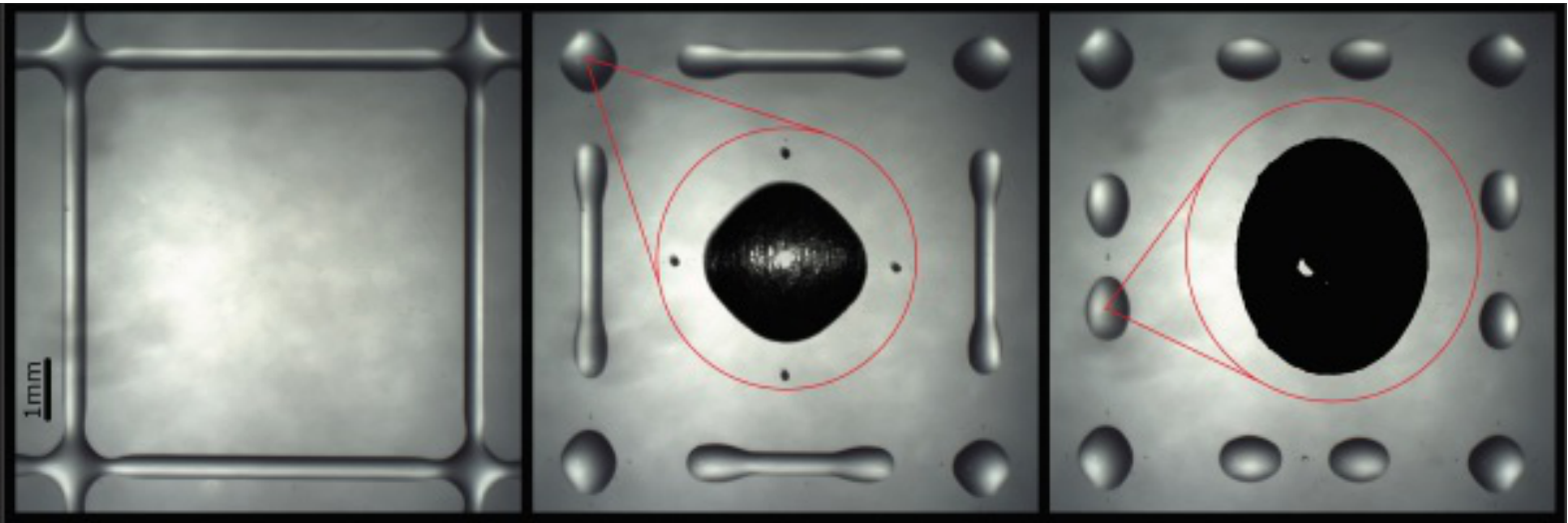}
    \caption{Time evolution of a square configuration obtained from two pairs of parallel filaments that are superimposed perpendicularly to each other. The insets show the two different types of drops that are formed at the intersections and along the filaments.}
    \label{fig:squares}
\end{figure}

A typical elementary unit of the grid is shown in Fig.~\ref{fig:squares}(a). After some seconds, four necks develop around the intersections till finally a drop at the crossings is formed as the necks break up. Then, the sides of the grids are now short filaments of an initial length, $L_i$, which start an axial dewetting process (see Fig.~\ref{fig:squares}(b)). Each one breaks up into a certain number of drops as shown in Fig.~\ref{fig:squares}(c). Interestingly, the drops resulting from the breakup of the filaments have a different morphology than those formed at the intersections (see the insets in Fig.~\ref{fig:squares}).

This arrangement of filaments has several advantages with respect to a single long filament. First, it enables us the possibility to study a new type of drop, namely, the one that is formed at the intersections of the captured filaments. We will see below that theses drops have features (shape, contact angle distribution, etc.) differing from those generated by the breakup of the short filaments (see insets in Fig.~\ref{fig:squares}). Second, since breakup processes mentioned above originate in the presence of ends or crossings in the filaments, we wish to study the relation between the length $L_i$ of the filaments and the number of resulting drops. Note that the short filaments formed after the breakups at the intersections acquire lengths and shaped ends that are very similar, allowing a convenient way to have simultaneously quasi identical unstable evolutions.
Consequently, we separate our analysis into two different aspects. One is the characterization of the drops at the intersections of the original long filaments and of those formed along the short filaments between the intersections (Section~\ref{sec:drops}). The other one is the rupture mechanism of these short filaments of controlled length, whose breakup leads to a linear array of drops (Section~\ref{sec:fil}). The compound effects of these two aspects lead to a bidimensional array of drops with different shapes and sizes that are arranged regularly. The possibility of a self-assembly of drops by using this processes can be of technological interest as explained in the introduction.

\section{Morphology of the sessile drops}
\label{sec:drops}

In these experiments we have two types of drops, those along the filaments, and those at the intersections. Both types of drops have non--circular footprints, but only the first type has been studied previously~\cite{gonzalez_07,rava_pof16}. The drops of the second type are different and have not been considered before. Therefore, we shortly revisit the approach done for the former, and extend the analysis to the latter.

The study of sessile drops with non--circular footprint is performed by looking for solutions of the equilibrium equation~\cite{rava_pof16},
\begin{equation}
 -\nabla^2 h + h =C,
\label{eq:p_eq1}
\end{equation}
where $h(\rho,\varphi)$ is the drop thickness, $\rho$, $\varphi$ are the radial and angular polar coordinates, respectively, and $C$ is a constant. Here, all lengths are expressed in units of the capillary length, $a$, and $0<\varphi<2\pi$. A solution of this equation can be written in the separable form $h= C + R(\rho)\Phi(\varphi)$. The two resulting uncoupled ordinary differential equations for $R$ and $\Phi$ can be solved straightforwardly, and we finally get a full solution of the form~\cite{rava_pof16}
\begin{equation}
h(\rho,\varphi)=C + \sum_{m=0}^\infty(A_m \cos m\varphi + B_m \sin m \varphi) J_m(\rho),
\label{eq:hpolar}
\end{equation}
after the diverging terms have been discarded. Here, $J_m(\rho)$ is the first kind Bessel function of order $m$, and $A_m$, $B_m$ are constants to be determined.

In the case of the drops along the filament axis, we can further use the fact that the drops have mirror symmetry respect to this axis as well as in the perpendicular direction. Thus, we must have both $B_m=0$ for all $m$, and $A_m=0$ for odd $m$. If we further assume that the shape of the drop can be reasonably estimated by the first four terms in the summation of Eq.~(\ref{eq:hpolar}) with even $m$, a simpler approximate expression can be used
\begin{equation}
h(\rho,\varphi)\approx C+A_0 J_0(\rho)+A_2 J_2(\rho) \cos 2 \varphi + A_4 J_4(\rho) \cos 4 \varphi + A_6 J_6(\rho) \cos 6 \varphi,
\label{eq:hdrop_fil}
\end{equation}
where the five unknown constants, $(C,A_0,A_2,A_4,A_6)$, must be determined from the experimental data. In fact, by measuring the values of the drop diameters $w_x$ and $w_y$, and the maximum thickness at its center $h_{max}$, we can form the following system of independent equations,
\begin{equation} 
 h \left( \frac {w_x}{2},0 \right)=0, \qquad h\left( \frac {w_y}{2},\frac {\pi}{2} \right)=0, \qquad h(0,0)=h_{max},
 \qquad \frac {\partial h}{\partial x}\left( \frac {w_x}{2},0 \right)=\theta_r, \qquad \frac {\partial h}{\partial y}\left( \frac {w_y}{2},\frac {\pi}{2} \right)=\theta_a,
 \label{eq:cond_dropfil}
\end{equation}
where the angles $\theta_a$ and $\theta_r$ stand for the (static) advancing and receding contact angles of the corresponding hysteresis cycle. These values are used because the contact line recedes along the filament ($\varphi=0$)  after the breakup, while it advances in the transverse direction ($\varphi=\pi/2$). The system of equations resulting from Eqs.~(\ref{eq:hdrop_fil}) and (\ref{eq:cond_dropfil}) can be solved analytically for $(C,A_0,A_2,A_4,A_6)$ with  which it is possible to calculate the shape of the footprint as well as the contact angle distribution around the drop periphery, $\theta(\varphi)$. The former can be measured from digitalized drop images, while the latter is obtained from the refraction pattern of the drop when impinged by a laser beam perpendicularly to the substrate~\cite{gonzalez_07,rava_pof16}. 

\begin{figure}[htb]
	\subfigure[]
        {\includegraphics[width=0.497\linewidth]{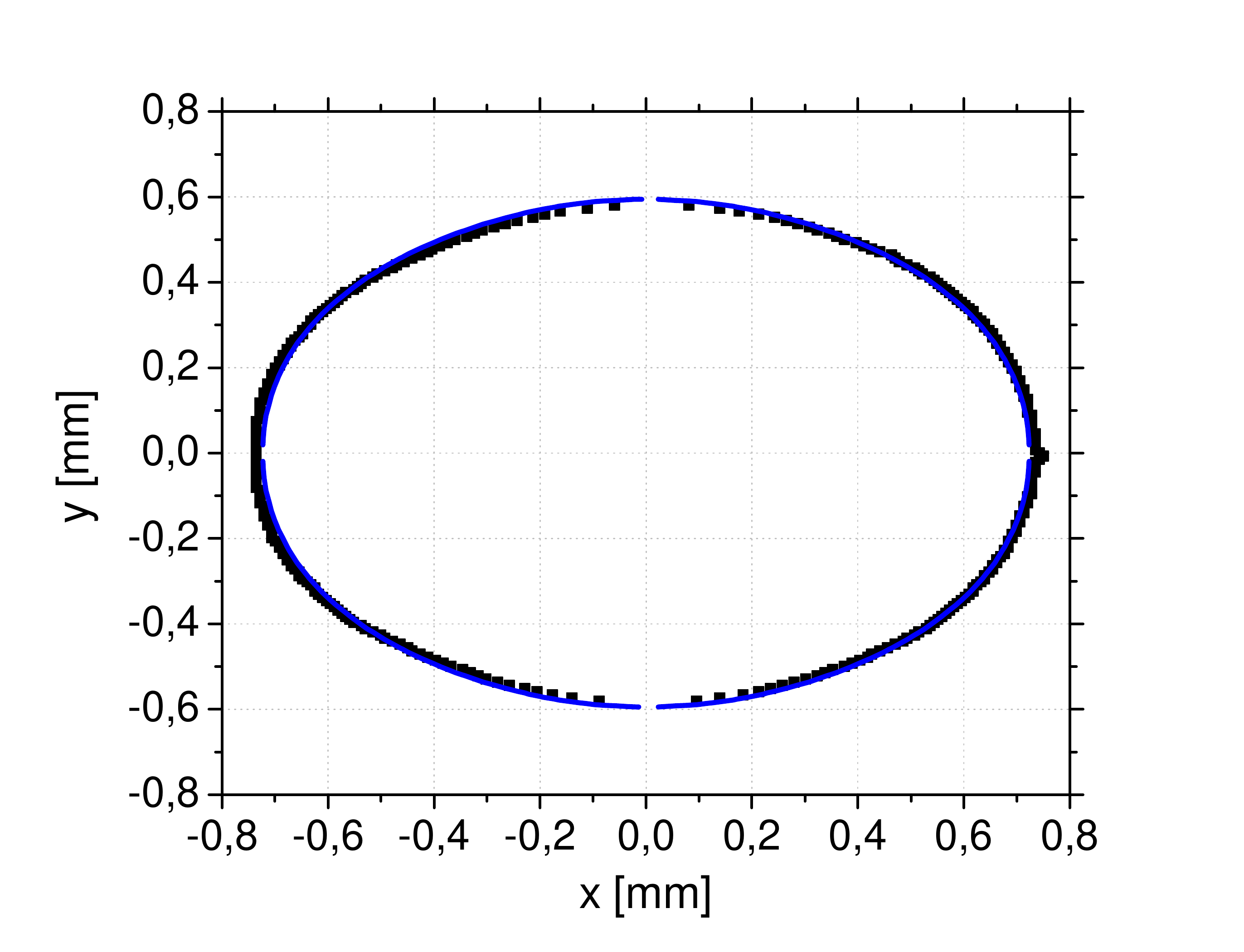}}
	\subfigure[]
        {\includegraphics[width=0.497\linewidth]{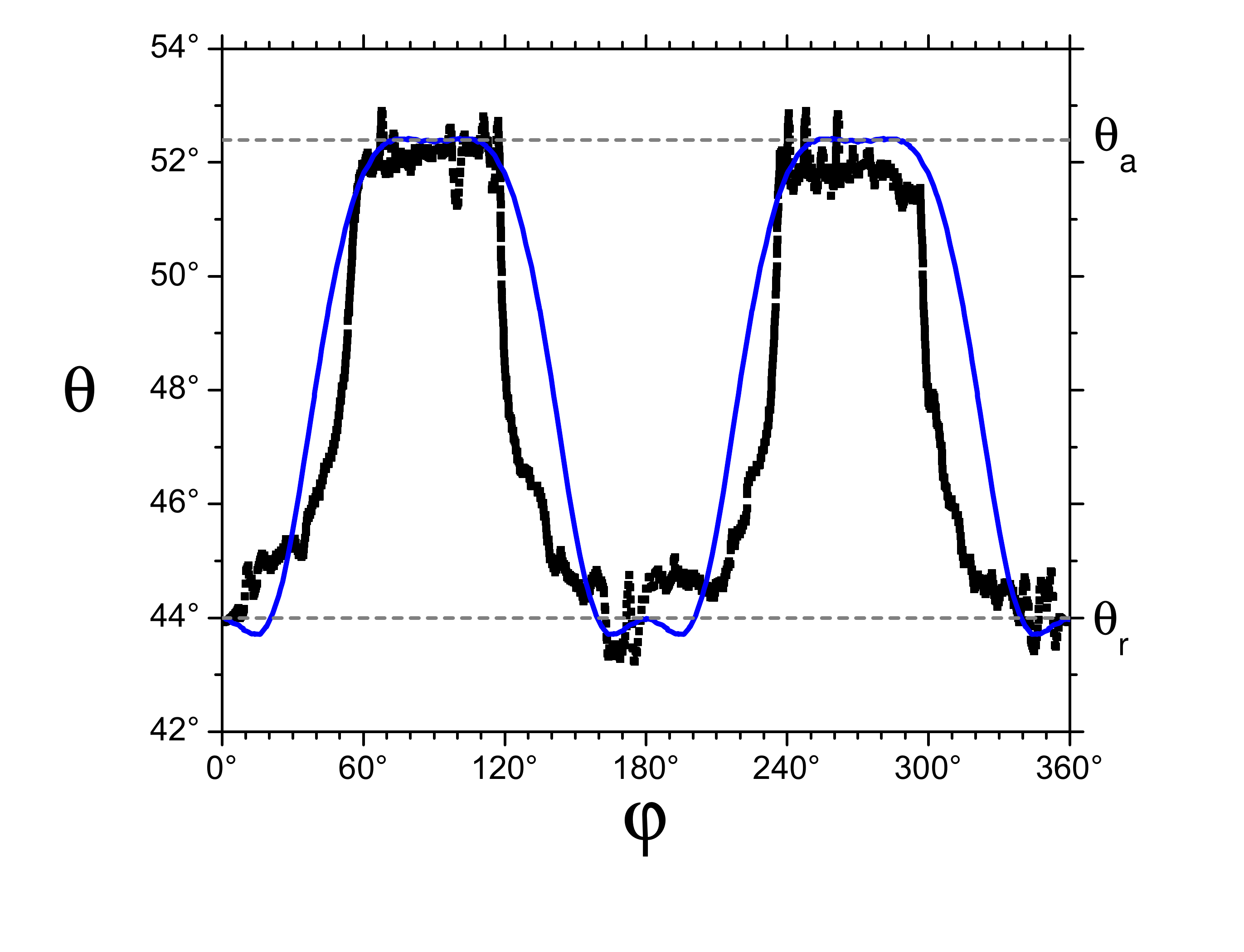}}
	\caption{{\it Drop at the filament}: (a) Non--circular footprint. (b) Contact angle at the drop periphery as a function of the azimuthal angle. The dashed lines ($\theta_a,\theta_r$) stand for the advancing and receding contact angles used in the calculation of the the theoretical curve. The symbols are the experimental data and the lines correspond to the theoretical model.}
	\label{fig:drop_fil}
\end{figure}

The comparison between the theoretical results and the experimental data for one drop along the filament is shown in Fig.~\ref{fig:drop_fil}. Clearly, the approximate solution is able to reproduce the quasi--elliptical shape of the footprint (see Fig.~\ref{fig:drop_fil}a). Furthermore, it gives account of the relation $\theta(\varphi)$ (see Fig.~\ref{fig:drop_fil}b), and it shows how $\theta$ changes from $\theta_r=44^\circ$ at $\varphi=0$ and $\varphi=\pi$ (along the filament axis) to $\theta_a=52.4^\circ$ at $\varphi=90^\circ$ and $270^\circ$ (in the transverse direction). However, the effect of the cut--off of terms in the series is more pronounced here than for the shape of the footprint. In fact, the comparison shows that the description of the rapid variation of $\theta(\varphi)$ at the borders of the plateau regions, say in the $\varphi$--intervals $(70^\circ,130^\circ)$ and $(240^\circ,300^\circ)$, requires more terms in the summation.

\begin{figure}[htb]
	\includegraphics[width=0.55\linewidth]{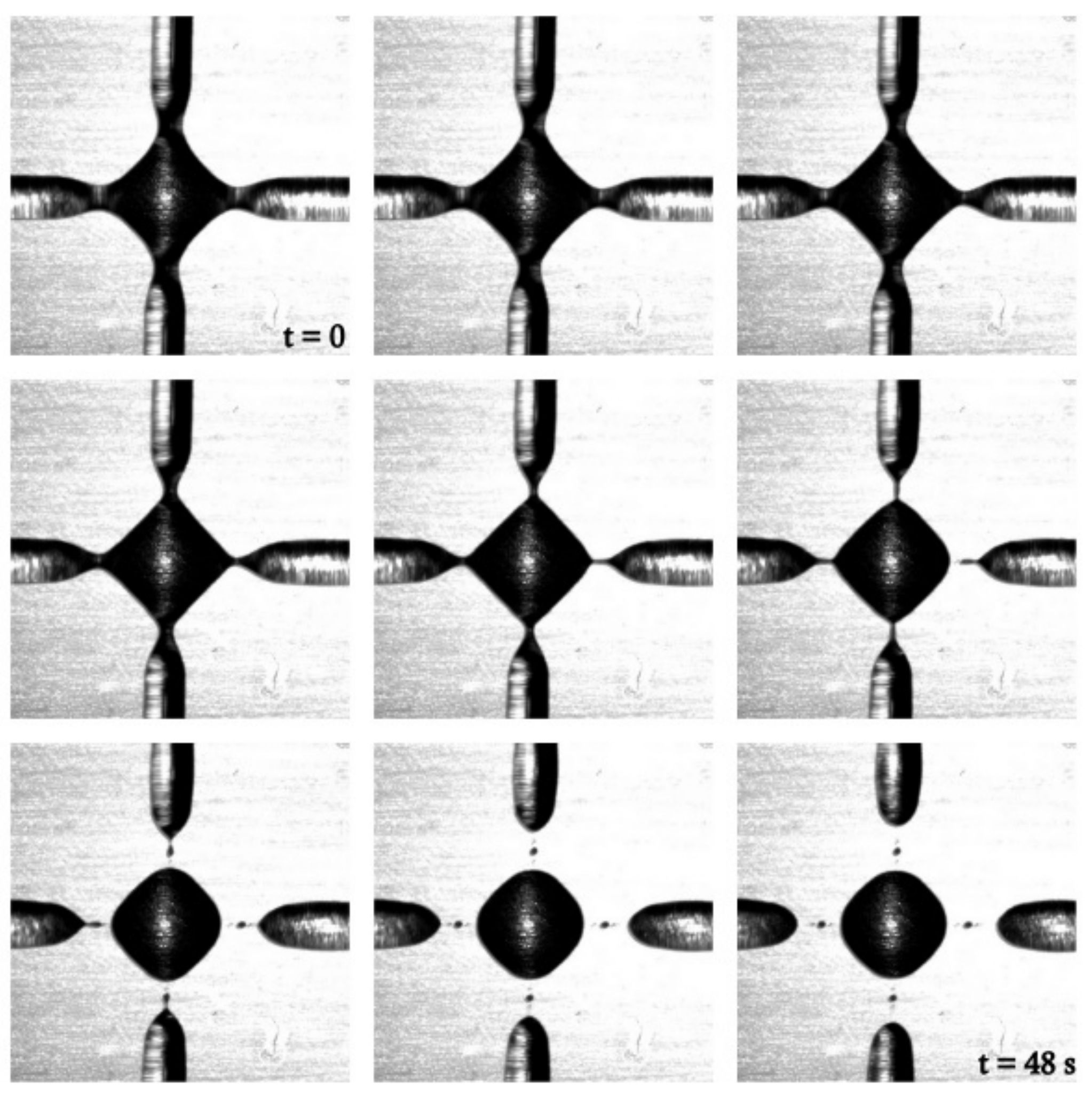}
	\caption{Time evolution of the drop formation at the intersection of two long filaments (time increases from left to right, and top to bottom).}
	\label{fig:cross}
\end{figure}

Unlike the drops along the filaments, in the case of drops at the intersections we have receding (dewetting) motions along both perpendicular filaments, while there are advancing (wetting) motions along the bisectors ($\varphi=\pm 45^\circ$ and $\pm 135^\circ$). In fact, as shown in Fig.~\ref{fig:cross},   after breakup there are regions in the contact line of the intersection which can be described as advancing straight and oblique lines. On the contrary, the vertex regions recede (dewet) along perpendicular directions. Interestingly, secondary droplets are also formed between the vertexes and the drops, whose study is out of the scope of this work.

Therefore, we must consider only solutions with biaxial symmetry, i.e. with multiplicity $4$, and write the following approximate expression of Eq.~(\ref{eq:hpolar}),
\begin{equation}
h(\rho,\phi)=C+ A_{0}J_{0}(\rho)+ A_{4}J_{4}(\rho)\cos(4\varphi)+  A_{8}J_{8}(\rho)\cos(8\varphi)
\label{eq:hdrop_int}
\end{equation}
with $(C,A_0,A_2,A_8)$ being four unknown constants. Analogously to the drops in the filament, these are determined from the boundary conditions
\begin{equation}
h\left( \frac {w_x}{2},0 \right)=0, \quad h(0,0)=h_{max}, \quad \frac{\partial h}{\partial x}\left(  \frac {w_x}{2},0 \right)= \theta_{r}^\prime, \quad \frac{\partial h}{\partial x}\left(  \frac {w_b}{2},\frac{\pi}{4}\right)= \theta_{a}^\prime,
\label{eq:cond_dropint}
\end{equation}
where $w_b$ is the drop width along the bisector, and the angles ($\theta_{r}^\prime,\theta_{a}^\prime$) are fitting parameters, which are expected to be close to  the values ($\theta_{r},\theta_{a}$) obtained from the filament drop. This system of equations for $(C,A_0,A_2,A_8)$ can also be solved analytically. Note that, unlike the conditions for the filament drop in Eq.~(\ref{eq:cond_dropfil}), we do not impose $h=0$ at $(w_b,\pi/4)$, since it is not necessary to determine the four coefficients in Eq.~(\ref{eq:hdrop_int}). It turns out that this condition is practically satisfied by the solution, since the resulting thickness there is less than $10^{-2} h_{max}$, so that the imposed slope at $(w_b,\pi/4)$ actually corresponds to a point on the contact line where $h \approx 0$.

\begin{figure}[htb]
	\subfigure[]
	{\includegraphics[width=0.497\linewidth]{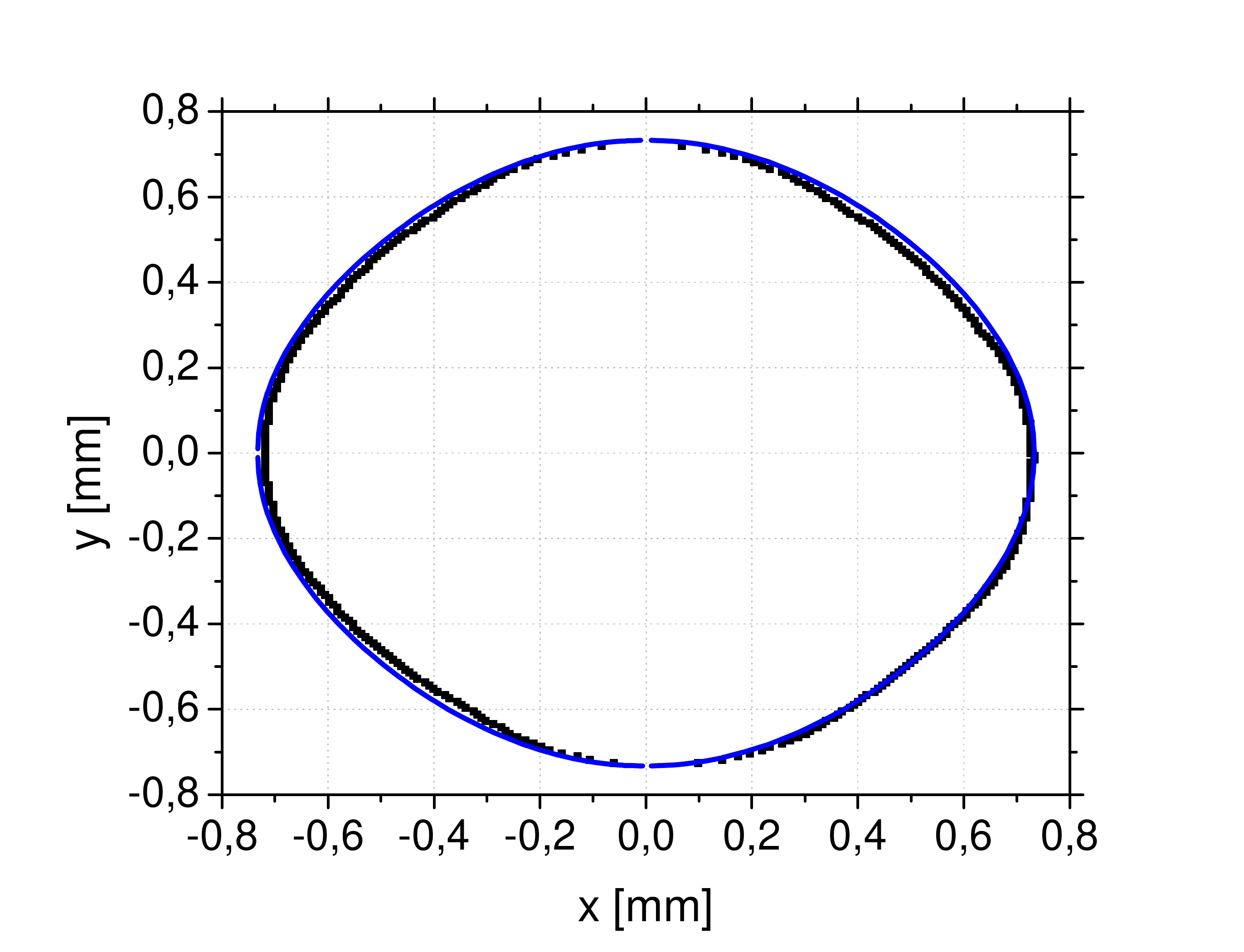}}
	\subfigure[]
	{\includegraphics[width=0.497\linewidth]{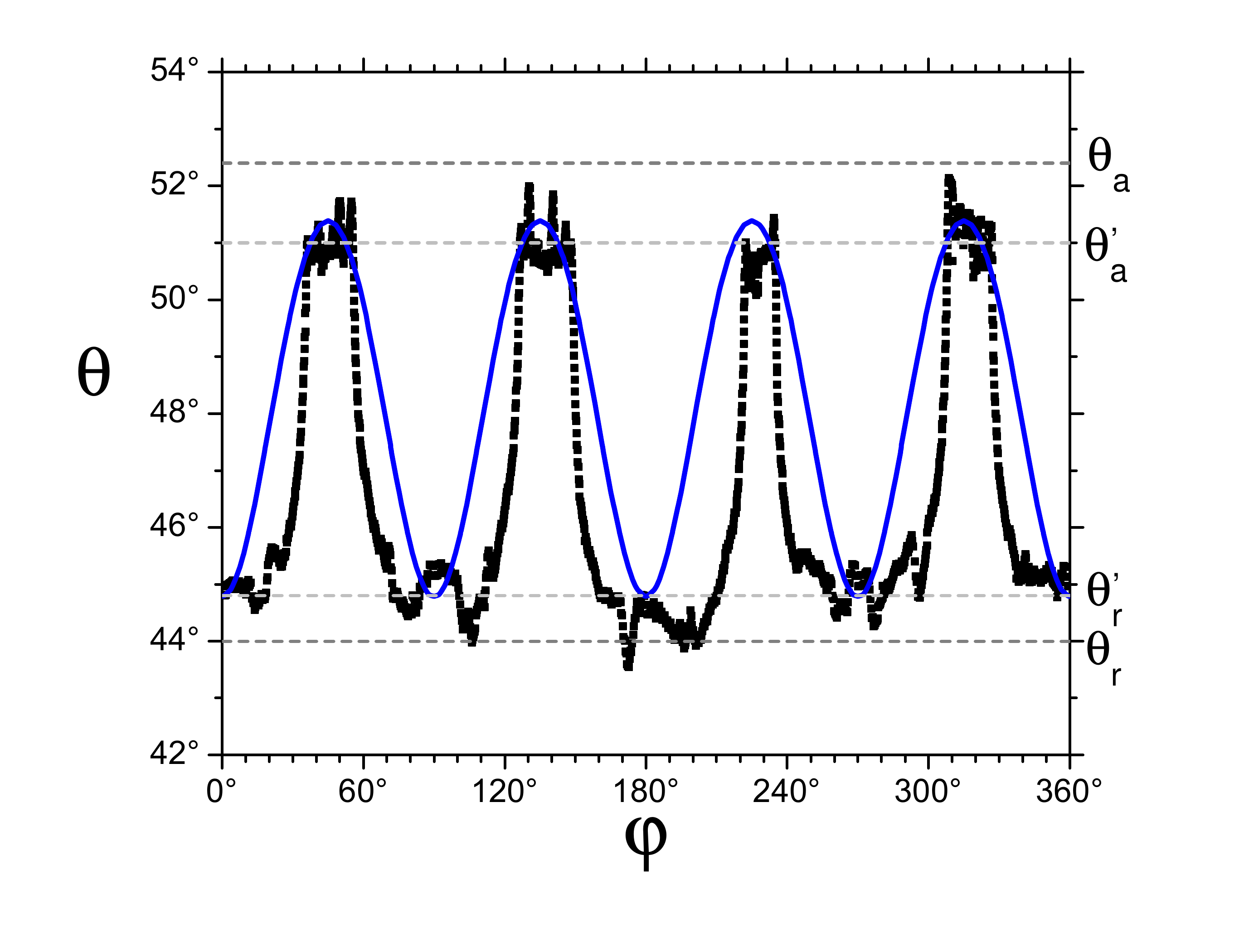}}
	\caption{{\it Drop at the intersection}: (a) Non--circular footprint. (b) Contact angle at the drop periphery as a function of the azimuthal angle. The dashed lines ($\theta_a^\prime,\theta_r^\prime$) stand for the advancing and receding contact angles used in the calculation of the the theoretical curve, and those for ($\theta_a,\theta_r$) correspond to the values obtained for the filament drop in Fig.~\ref{fig:drop_fil}b. The symbols are the experimental data and the lines correspond to the theoretical model.}
	\label{fig:drop_int}
\end{figure}

A comparison of these theoretical results with the experimental data is shown in Fig.~\ref{fig:drop_int}.  We observe that for $\theta_{r}^\prime=44.8^\circ$ and $\theta_{a}^\prime=51^\circ$, we have a very similar degree of agreement between theory and experiment for both the footprint shape and the angular distribution, $\theta(\varphi)$, as obtained for the filament drop case in Fig.~\ref{fig:drop_fil}. The main difference between the footprints of both types of drops is that now it adopts a quasi--square shape (Fig.~\ref{fig:drop_int}a), instead of an ellipsoidal one. On the other hand,  the angular distribution in Fig.~\ref{fig:drop_int}b has four maximums and four minimums for $\theta(\varphi)$. As expected, the latter are located at the middle of the sizes of the rounded square ($\varphi=\pm 45^\circ$ and $\pm 135^\circ$), and the former at its vertexes ($\varphi= k\, 90^\circ$, $k=0,1,2,3$). Interestingly, the $\theta$--interval for the intersection drop, namely ($44.8^\circ,51^\circ$), lies inside the one for the filament drop, ($44^\circ,52.4^\circ$), which is consistent with the expected hysteretic behavior of the contact angle.

\section{Rupture of short filaments}
\label{sec:fil}

\begin{figure}[htb]
	\includegraphics[width=\textwidth]{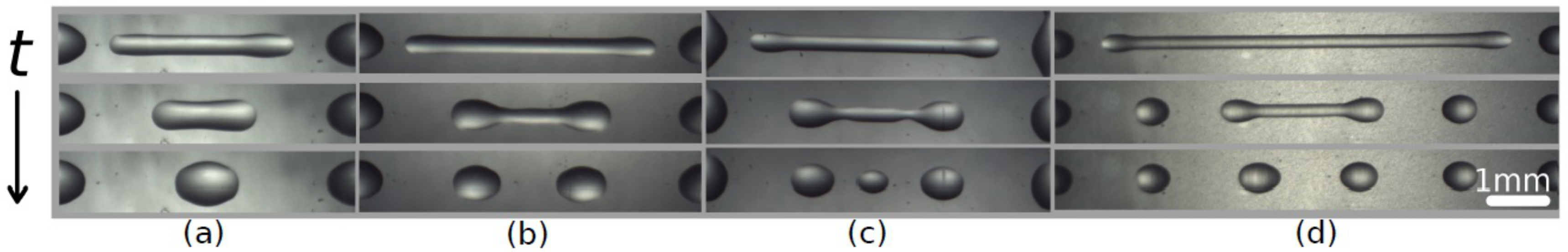}
	\caption{Three times of the evolution of four filaments with different aspect ratios, $\Delta=L_i/w$: (a) $\Delta=7.34$, (b) $\Delta=13.5$, (c) $\Delta=17.84$, and (d) $\Delta=34.65$. The upper, middle, and bottom rows correspond to initial ($t=0$), intermediate ($t=110$, $130$, $200$, $210$~s) and final ($t=210$, $240$, $290$, $360$~s) times of the evolution, respectively.}
	\label{fig:4fil}
\end{figure}

The time evolution of the contact line profiles are shown in Fig.~\ref{fig:4fil} for some filaments seen from top. In this study, we carry out experiments with $L_i$ and $w$ in the ranges $2$~mm $<L_i <8$~mm  and $0.25$~mm $< w <0.45$~mm. As expected, the number of drops resulting from the breakup process occurring between the end heads strongly depends on $L_i$ or, more precisely, on the aspect ratio $\Delta=L_i/w$. Our experimental device allows us to explore the range $5 <\Delta<38$.

\begin{figure}[htb]
\includegraphics[scale=0.4]{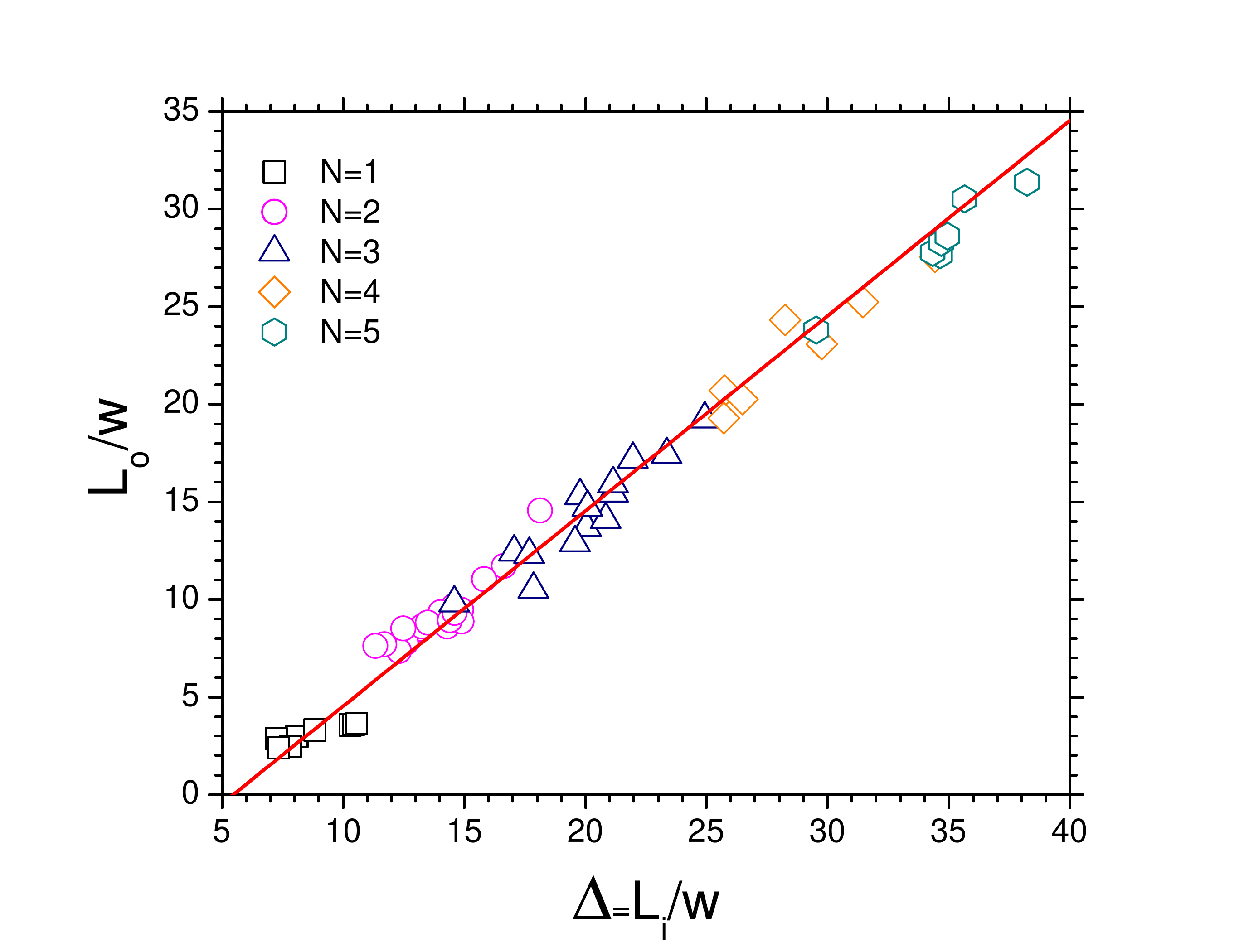}
\caption{Filament length when the heads have stopped, $L_0$, versus the initial length, $L_i$, in units of the filament width, $w$. The different symbols stand for the resulting number of drops. The line corresponds to  $L_0/w = L_i/w-(5.46\pm 0.13)$}
\label{fig:LiL0}
\end{figure}

When both heads have stopped, their tips have receded a certain distance $L_d$, so that the new filament length is 
\begin{equation}
 L_0=L_i-2 L_d.
 \label{eq:L0Li}
\end{equation}
In Fig.~\ref{fig:LiL0} we show by symbols the measured values of $L_0$ versus $L_i$ in units of the corresponding filament width, $w$. The distinctive  symbols correspond to different numbers of drops that result from the filament breakups. By approximating these data with Eq.~(\ref{eq:L0Li}) we find $L_d=(2.73\pm 0.06) w$, which is consistent with the results in~\cite{rava_pre17}, where the proportionality between $L_d$ and $w$ was predicted. The  axial displacement of the contact line after the first neck breakup is important to estimate the final number of drops, since it is the ratio $L_i/w$ what actually determines it.

\section{Model for the neck formation behind the head}
\label{sec:model}

Here, we develop a simple hydrodynamic model to account for the process formation of the neck at a certain distance away from the head. We will take into consideration the approximate shape of the head when it stops receding, which is close to the moment when the neck begins to form. As a result of the characteristic positions of the resulting breakups and the sizes of the drops given by the model, we obtain the predicted number of drops formed from a filament of given length and width, $w$, which can be compared with the experimental evidence. 

\begin{figure}[htb]
\includegraphics[width=0.45\linewidth]{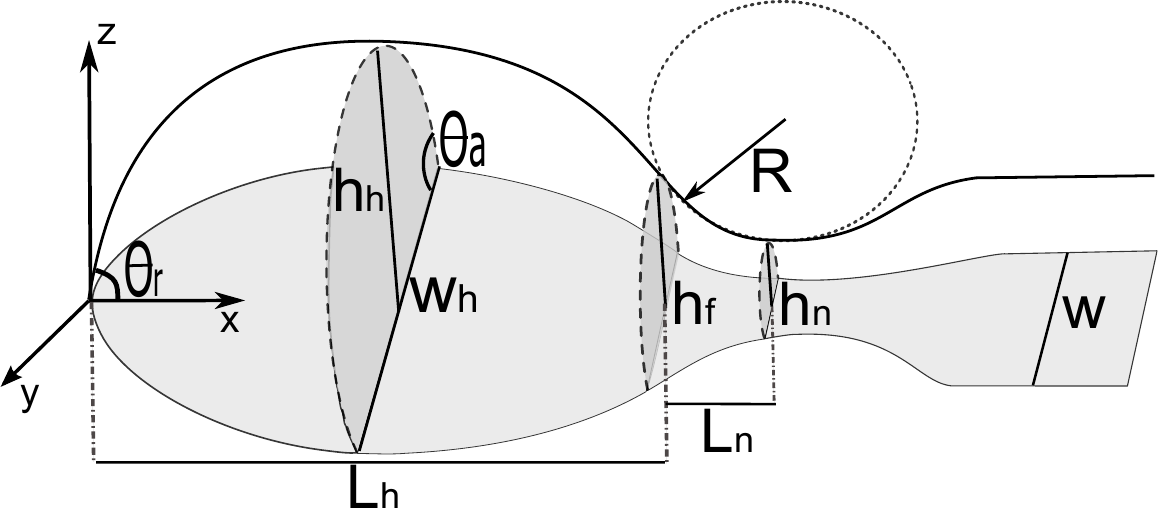}
\caption{Sketch of the head and neck regions showing the parameters used in the model.}
\label{fig:sketch_fil2drops}
\end{figure}

We consider that the initial condition, $t=0$, is given by the filament formed after the breakup from the intersections, and that its length is $L_i$. For $t>0$, the ends of the filament recede along the axis of the filament, and a head starts growing in that region (see Fig.~\ref{fig:4fil}), and later on, the motion of the tip stops with a filament length equal to $L_0$. After a while, a neck region starts forming in the filament some distance away from the head. Our aim is to model the process resulting from the flow that develops between the head and the filament, which determines the actual position of a neck, where the width is minimum (see Fig.~\ref{fig:sketch_fil2drops}). Since the experiments show that the breakup processes occur as a consequence of the unstable pinch off of the necks, we posit that the determination of their positions is closely related to the number and size of drops that will result from a given filament.

Once the axially dewetting motion of the ends has finished, the pressure in the head, $p_h$, is balanced by that in the connecting region with the filament, $p_f$. Assuming that the filament cross section in that region is also circular and that the contact lines are ready to dewet there, we have 
\begin{equation}
p_f= \frac{2 \gamma \sin \theta_r}{w}.
\label{eq:pf}
\end{equation}
On the other hand, considering the axial and transverse curvature radii at the head region, the pressure in this region can be estimated as
\begin{equation}
 p_h= 2 \gamma \left( \frac{\sin \theta _a}{w_h} + \frac{\sin \theta_r}{L_h}\right),
 \label{eq:pb}
\end{equation}
where $w_h$ and $L_h$ are the width and length of the head, respectively. Now, we assume that the shape of the head (or at least the region between the apex and the tip) does not differ significantly from that of the resulting drop at rest, a fact that has been observed in the experiments. Thus, we can use the following geometrical property
\begin{equation}
 L_h =\frac{\theta_a}{\theta_r} w_h,
 \label{eq:Lh}
\end{equation}
that has been found to be valid for drops in similar experiments with long filaments from both experimental and theoretical grounds~\cite{rava_pof16}.
As a consequence, due to the pressure balance $p_h=p_f$, we find the following expression for the width of the head as a function of the contact angles and the width of the filament:
\begin{equation}
 w_h= w \frac{\theta_a \sin \theta_a+\theta_r \sin \theta_r }{ \theta_a \sin \theta_r}.
 \label{eq:wh}
\end{equation}
In our case, we have from Eqs.~(\ref{eq:Lh})-(\ref{eq:wh}) that
\begin{equation}
 L_h= 2.34 w, \qquad w_h=1.98 w.
 \label{eq:Lh_wh}
\end{equation}

As the neck becomes thinner at a certain distance from the head, the pressure there increases and, consequently, a fluid motion away from the neck is established (see Fig.~\ref{fig:sketch_fil2drops}). We assume that this flow is of the Stokes type, so that there is a balance between the gradients of pressure and viscous stress,
\begin{equation}
\nabla p = \mu \nabla ^2 v.
\label{eq:stokes}
\end{equation}
Here, we consider this balance between the neck region and the filament by means of the approximation,
\begin{equation}
\frac{p_f -p_n}{L_n}= \mu \frac{v}{h_f^2},
\label{eq:stokes_app}
\end{equation}
where $p_n$ is the pressure at the neck, 
\begin{equation}
 h_f= w \frac{1-\cos \theta_r}{2 \sin \theta_r}
 \label{eq:hf}
\end{equation}
is the thickness of the filament at the region connecting the head with neck (see Fig.~\ref{fig:sketch_fil2drops}), and $v$ is the mean axial flow velocity. Here, the value of $p_n$ can be estimated as 
\begin{equation}
 p_n=\gamma \left( \frac{2 \sin \theta_r}{w_n}- \frac{1}{R} \right),
 \label{eq:pn}
\end{equation}
where $R$ is the axial radius of curvature in that region. Since the thickness at the filament, $h_f$, and at the neck, $h_n$, do not differ too much ($h_f-h_n \ll L_n$), we can approximate this radius by
\begin{equation}
 R= \frac{L_n^2}{2(h_f- h_n)},
 \label{eq:Rcurv}
\end{equation}
where $h_n$ is obtained under the assumption of a neck with circular cross section,
\begin{equation}
 h_n= w_n \frac{1-\cos \theta_r}{2 \sin\theta_r},
 \label{eq:hn}
\end{equation}
being $w_n$ the neck width.

\begin{figure}[htb]
	\includegraphics[width=0.7\textwidth]{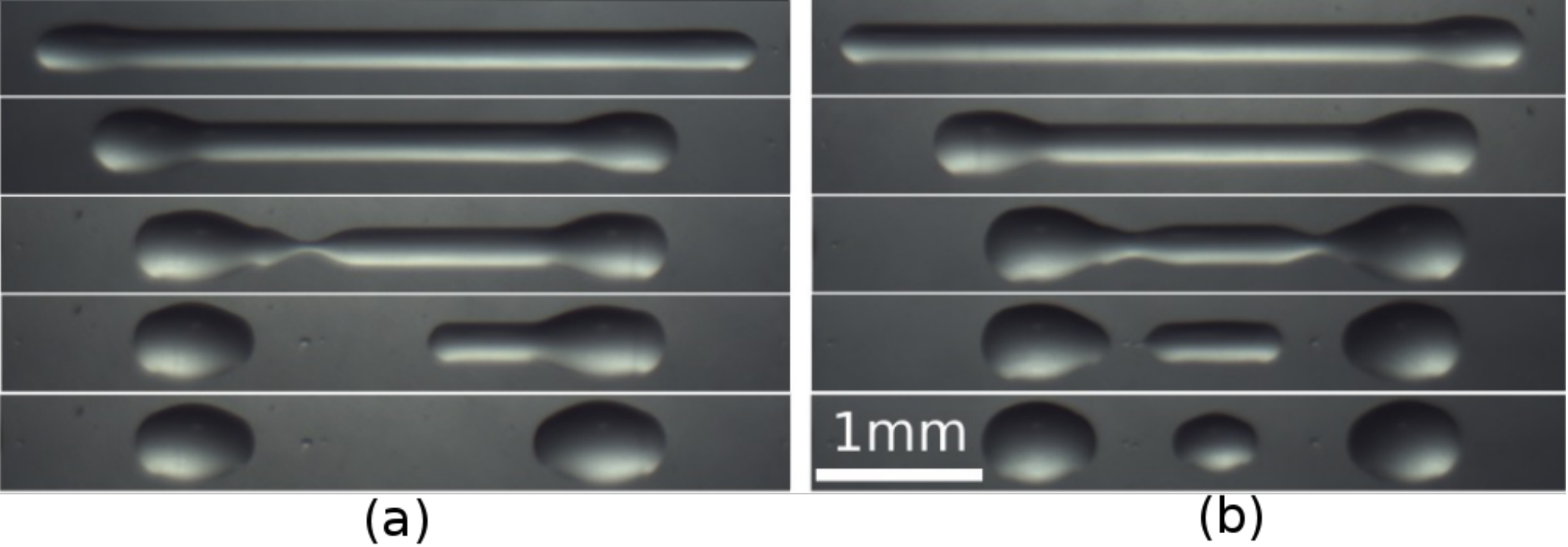}
	\caption{Time evolution of two filaments practically identical which break up into a different number of drops: (a) two drops ($L_i=5.44$~mm and $w=0.33$~mm, $\Delta=16.5$), and (b) three drops ($L_i=5.66$~mm and $w=0.32$~mm, $\Delta=17.7$).}
	\label{fig:Li2-3}
\end{figure}

On the other hand, the mean velocity $v$ can be written as $v=\omega L_n$, where $\omega $ is the maximum growth rate of the linear stability analysis of an infinitely long filament. For the viscous regime, we have~\cite{brochard_lang92}
\begin{equation}
 \omega=   \frac{0.379 \gamma}{30 \mu} \frac{\thetạ_r^3}{w}.
 \label{eq:omega}
\end{equation}
Thus, at a time $\tau=1/\omega$, we have $w_n=w/e$. By using Eqs.~(\ref{eq:pf})-(\ref{eq:omega}) along with the numerical values of the parameters, we finally obtain an expression for the Stokes balance in Eq.~(\ref{eq:stokes_app}) as a biquadratic equation for $L_n$ in terms of the width filament, $w$,
\begin{equation}
L_n^2 + \frac{8.496 w^4}{L_n^2} = 24.81 w^2.
\end{equation}
In consequence, we have two possible values for $L_n$, which we call `short' and `long', 
\begin{equation}
 L_s= 0.59 \, w, \qquad L_l= 4.95 \, w.
 \label{eq:Ls_Ll}
\end{equation}
An experimental example of these two solutions is shown in Fig.~\ref{fig:Li2-3}a, where the formation of the neck behind the left head is clearly visible. Therefore, the number of drops that result from a given filament depends on how many necks 
can be formed when the filament has reached the length $L_0$ after having started with $L_i$. 
It is interesting to note that
\begin{equation}
L_l\approx L_d+L_h=(5.07 \pm 0.06)\,w.
\label{eq:LlLdLh}
\end{equation}
This means that the long length corresponding to a breakup allows for the formation of a head of size $L_h$ after a retraction close to $L_d$, which is precisely the receding distance observed at the ends of the filaments.  This fact has been observed in several experiments (see e.g. Fig.~\ref{fig:Li2-3}a). Equation~(\ref{eq:LlLdLh}) will be useful in understanding the relationship between $L_i$ and the number of resulting drops, $N$, as it will be explained in the next section. 

Note that very small secondary drops~\cite{gonzalez_07} can be observed in the region of the breakup (see Fig.~\ref{fig:Li2-3}a). These secondary drops are generally at a distance close to $L_s$ from one of the bulges. Their origin and behavior is different from the primary drops we are interested in, but their presence at $L_s$ is indicative of where necks have occurred and that $L_s$ is a relevant distance for explaining the ruptures.

\section{Filament aspect ratio versus number of drops}

Based on the previous models and analysis, we establish here the conditions that must be fulfilled to obtain a given number of drops from the rupture of a filament of certain initial length, $L_i$. In the following, we perform the corresponding analysis as the number of drops increases.

\begin{figure}[htb]
\includegraphics[width=0.6\linewidth]{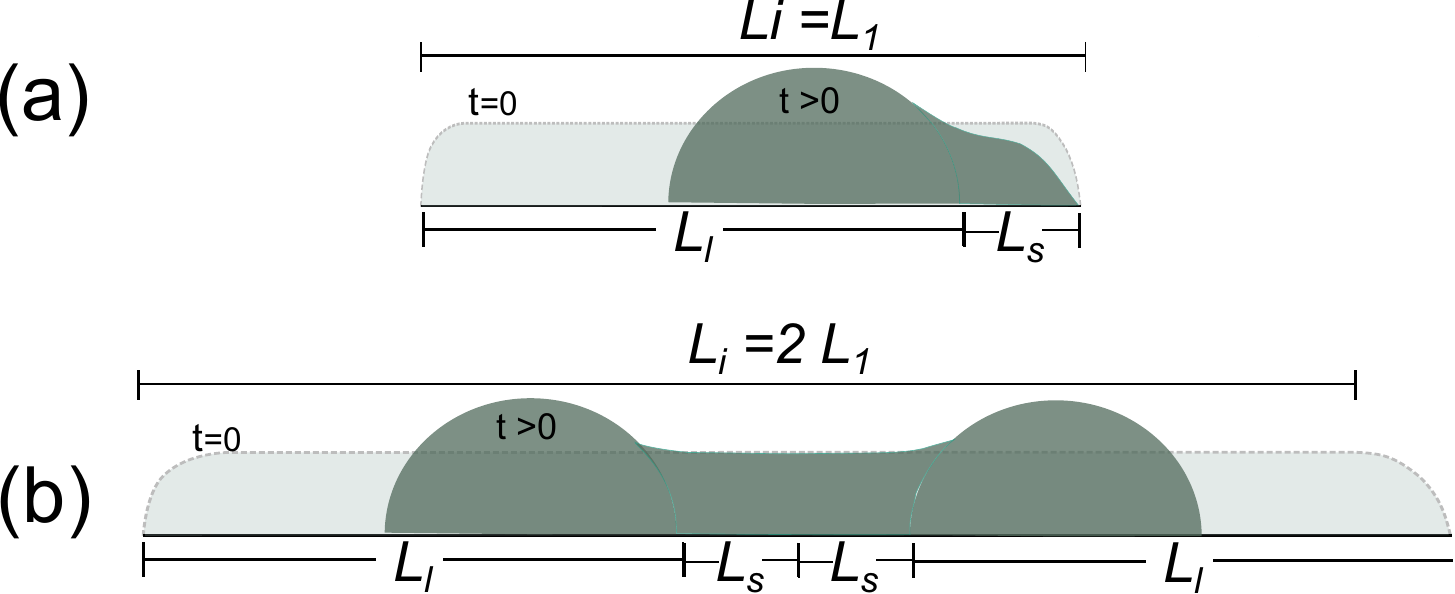}
\caption{Sketches of the filament showing the parameters used in the model to define the limiting lengths of the filaments that yield: (a) one drop, and (b) two drops.}
\label{fig:sketch}
\end{figure}

\begin{figure}[htb]
	\includegraphics[width=0.7\textwidth]{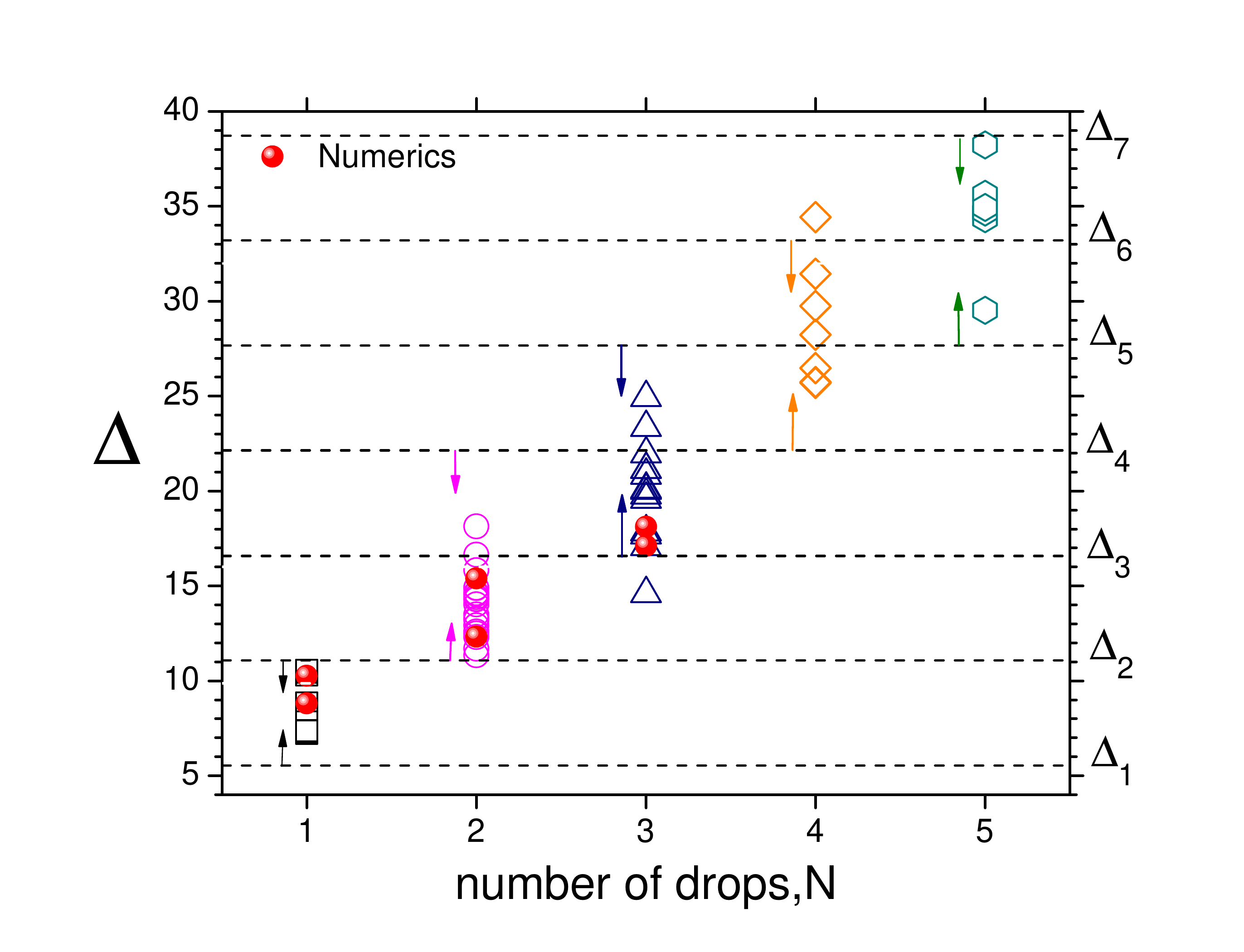}
	\caption{Number of drops, $N$, versus different aspect ratios, $\Delta=L_i/w$. The horizontal lines correspond to the limiting lengths predicted by the analysis, $\Delta_N=L_N/w$, and the solid circles to the numerical simulations.}
	\label{fig:ndrops}
\end{figure}

Let us first consider the conditions for the appearance of a single drop. Of course, a very short filament will retract to one drop. However, not all of these filaments allow for the formation of large enough drops where a simultaneous wetting of previously dry regions in the transverse direction to the filament occurs while the dewetting process is proceeding axially. Since we are interested in relatively large drops that are similar to those appearing in very long filaments, it is legitimate to pose the question of which is the minimal length leading to the formation of a single anisotropic drop comparable to those observed in long filaments. In order to obtain this type of drops, one needs a retraction distance leading to the formation of a head with axial length $L_h$. Therefore,  $L_i$ could not be less than $L_l\approx L_d+L_h$ plus a short tail on the other side of the head (see Fig.~\ref{fig:sketch}(a)). If the possible formation of a breakup leading to another drop is to be avoided, then this tail cannot be larger than $L_s$. Consequently, the minimal length to form a single drop of this kind is
\begin{equation}
L_1 \equiv L_l+L_s =5.54\, w,
\label{eq:L1}
\end{equation}
as shown by the lowest horizontal line in Fig.~\ref{fig:ndrops}. 

If $L_i= L_2 \equiv 2 L_1$, the possibility of forming two drops cannot be prevented since a breakup in the small bridge between the two heads formed from both ends of the filament is long enough to allow for the formation of a breakup at a distance $L_s$ from each head (see Fig.~\ref{fig:sketch}(b)). Following a similar reasoning, a general formula for the limit of $L_i$ to allow the formation of $N$ drops can be written as:
\begin{equation}
L_N \equiv N L_1,
\label{eq:Ln}
\end{equation}
where $N=1,2,...$. In Fig.~\ref{fig:ndrops}, we compare this prediction with the experimental lengths, $L_i$, that give place to a certain number of drops. Considering the approximations made in the model, we observe a very good agreement between experiment and the theoretical model.

Note however that these limits are only lower limits for the existence of a certain number of drops, but not the upper ones. For example, when $L_i$ is slightly below $L_2$ there is the possibility that both heads coalesce into a single drop. Then, the upper limit of one drop can be reasonably be estimated as $L_2$. Regarding the upper limit for more drops, this coalescence process could occur on both sides of the remaining bridge, and therefore its maximum length should be $2 L_1$. Then, the upper limit for the formation of $N$ drops can be written as $(N+2) L_1$ for $N\geq 2$. This means that the upper limit for more than two drops is coincident with the lower limit for $(N+2)$ drops. Even though the model is based on some rough approximations, these predicted limits agree very well with the experimental data (see Fig.~\ref{fig:ndrops}).

\begin{figure}[htb]
	\includegraphics[width=0.5\textwidth]{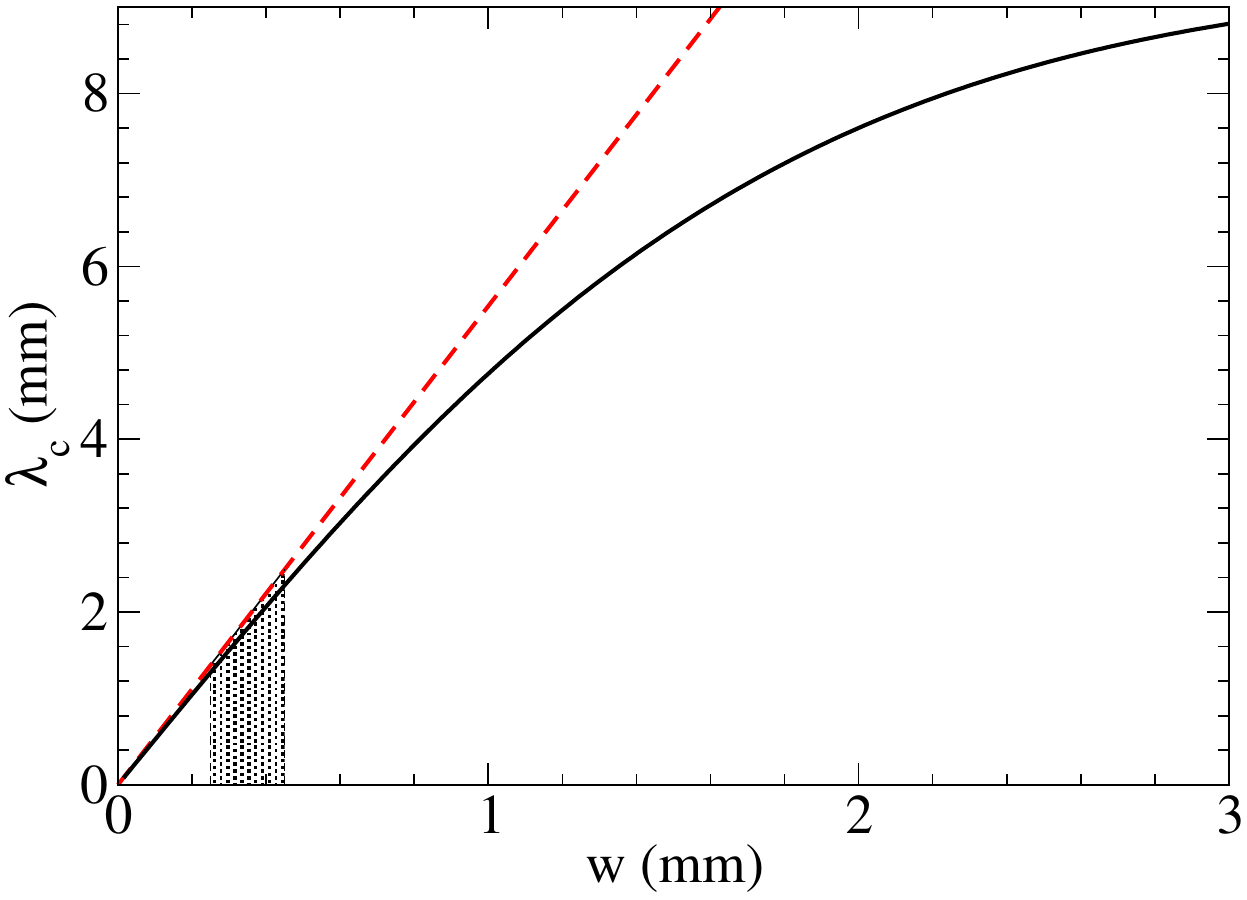}
	\caption{Critical (marginal) wavelength of the linear stability analysis for an infinitely long filament, $\lambda_c$, versus the filament width, $w$. The dashed straight line corresponds to the linear dependence of $L_1$ with $w$, as given by Eq.~(\ref{eq:L1}). The experimental interval of the filament widths, $(0.25,0.45)$~mm, is shown as a hatched region.}
	\label{fig:lc}
\end{figure}

The characteristic length $L_1$ can be compared with the critical (marginal) wavelength, $\lambda_c$, of the linear stability analysis for an infinitely long filament, given by~\cite{sekimoto_annphys87,dgk_pof12}
\begin{equation}
 s \tanh(q_c s) \tanh(s)=1,
 \label{eq:qc}
\end{equation}
where $s=w/a$, $q_c=2\pi a/\lambda_c$, and $a=\sqrt{\gamma/\rho g}$ is the capillary length. The resulting dependence between $\lambda_c$ and $w$ is shown in Fig.~\ref{fig:lc}, where we also plot the linear relationship between $L_1$ and $w$ as given by Eq.~(\ref{eq:L1}). Clearly,  Eq.~(\ref{eq:qc}) can be accurately approximated by this straight line within the range of our experiments, namely $w \in (0.25,0.45)$~mm. 

This comparison shows that the critical wavelength of the instability predicted for an infinite filament corresponds to the observations and the model characteristic lengths up to filaments as short as those which give place to a single drop. Interestingly, while both models for finite and infinite filaments yield similar lengths in the range of parameters from our experiments, we consider successive steps in the breakup process as actually seen in the experiments, while the infinite filament theory predicts simultaneous breakups, which have not been observed. 

\section{Numerical simulations of the evolution of short filaments}
\label{sec:num}
In this section we numerically simulate the evolution of a filament of length, $L_i$, and width, $w$. We consider that at $t=0$ the filament has a body shape of a cylindrical cap of length $L_{cyl}<L_i$, width $w$, and transverse equilibrium contact angle, $\theta_a$, along both parallel contact lines. In order to emulate the ends of an actual filament, which have rounded shapes due to the breakup process that took place at the intersections for $t<0$, we approximate them by additional ellipsoidal caps at the ends of the cylinder ($x=0$ and $x=L_{cyl}$). 

Thus, the fluid domain is composed by a cylindrical cap plus two ellipsoidal caps at the ends, and the whole filament has length $L_i$. The ellipsoidal caps are determined by two parameters, namely, the maximum width in the transverse direction, $w_c$, and the contact angle at the end, $\theta_x$. In all the cases, we take $\theta_x=25^\circ$ since this is the contact angle observed in the experiments just after the breakup~\cite{rava_pof16}, and $w_c \gtrsim w$ as necessary to better adjust the shape of the initial head to the actual initial condition of each experiment.

The time evolution of this liquid filament is obtained by numerically solving the dimensionless Navier-Stokes equation
\begin{equation}
La \left[ \frac{\partial \vec {v}}{\partial t}+(\vec{v} \cdot \vec {\nabla} )  \vec{v} \right] 
= - \vec {\nabla} p + \nabla ^{2} \vec {v} -\vec{z},
\label{eq:NS}
\end{equation}
where the last term stands for the gravity force. Here, the scales for the position $\vec x=(x,y,z)$, time $t$, velocity $\vec {v}=(u,v,w)$, and pressure $p$ are the capillary length $a$, $t_c=\mu a/\gamma$,$\gamma/\mu$, and $\gamma/a$, respectively. Therefore, the Laplace number is $La=\rho \gamma a/\mu^2$. In our experiments we have $a=1.49$ mm and $La=0.006$, so that inertial effects are practically irrelevant. The $x$ and $y$--axes are assigned along and across the original filament, respectively.  Besides, the normal stress at the free surface accounts for the Laplace pressure in the form
\begin{equation}
 \Sigma_n = -\left( \vec \nabla_{\tau} \cdot \hat n \right) \hat n,
\end{equation}
where $\hat n=(n_x,n_y,n_z)$ and $\hat \tau$ are the versors standing for the normal and tangential directions to the free surface. Since, the surrounding fluid (e.g., air) is passive, we assume that the tangential stress is zero at this surface, i.e. $\Sigma_{\tau}=0$. 

As regards to the boundary condition at the contact line, the dynamic contact angle, $\theta$, is given by the dimensionless contact line velocity, $Ca=\mu v_{cl}/\gamma$, according to the hybrid model~\cite{rava_pre17}
\begin{equation}
 \theta^3 = \arccos^3 \left[ \cos \theta_0 - \frac{1}{\Gamma} \sinh^{-1} \left( \frac{Ca}{Ca_0} \right) \right] + 
 9 Ca \ln \left( \frac{1}{\hat \ell} \right),
 \label{eq:theta_fit_num}
\end{equation}
where $Ca_0=\mu v_0/\gamma$, and $\hat \ell = \ell/a$. The contact line velocity is calculated from the velocity field as $Ca= N_x u + N_y v$, where $(N_x,N_y)=(n_x,n_y)/\sqrt{n_x^2+n_y^2}$ is the versor normal to the contact line. Note that this condition introduces a high nonlinearity to the problem, since the solution itself, namely the velocity field at $z=0$, yields the corresponding contact angle. Since we are using the same type of fluid (PDMS) and substrate as in~\cite{rava_pre17}, the values of the coefficients $\Gamma$, $\ell$, and $v_0$ are equal to those given in that reference, i.e. $\Gamma= 95.4553$, $\ell= 0.0008302 a=1.24 \times 10^{-4}$~cm, and $v_0=6.2121 \times 10^{-7}$~cm/s.

We use a Finite Element technique in a domain which deforms with the moving fluid interface by using the Arbitrary Lagrangian-Eulerian (ALE) formulation~\cite{hughes_cmame81,donea_cmame82,christodoulou_cmame92,hirt_cmame97}. The interface displacement is smoothly propagated throughout the domain mesh using the Winslow smoothing algorithm~\cite{winslow_jcp66,knupp_ec99}. The main advantage of this technique is that the fluid interface is and remains sharp~\cite{tezduyar_cmame06}, while its main drawback is that the mesh connectivity must remain the same, which precludes achieving situations with a topology change (e.g., when the filament breaks up). The default mesh used throughout is unstructured, and typically has $3\times 10^4$ triangular elements (linear elements for both velocity and pressure).  The mesh nodes are constrained to the plane of the boundary they belong to except those at the free surface.

\begin{figure}[htb]
	\subfigure[$L_i=0.361$~cm, $w=0.037$~cm ($\Delta=9.76$)]
        {\includegraphics[width=1.0\linewidth]{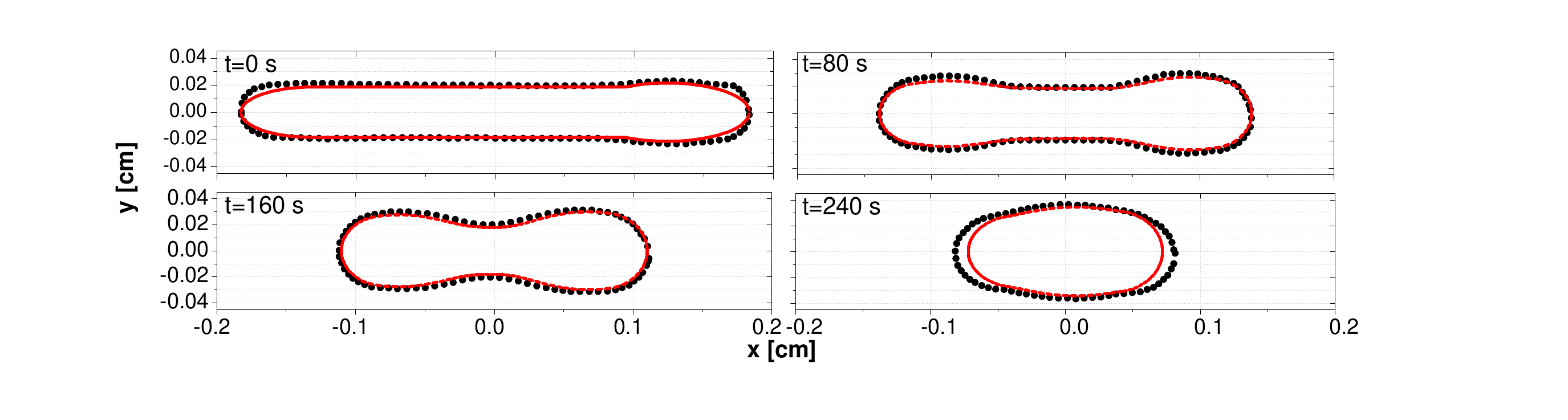}}
	\subfigure[$L_i=0.441$~cm, $w=0.036$~cm ($\Delta=12.25$)]
        {\includegraphics[width=1.0\linewidth]{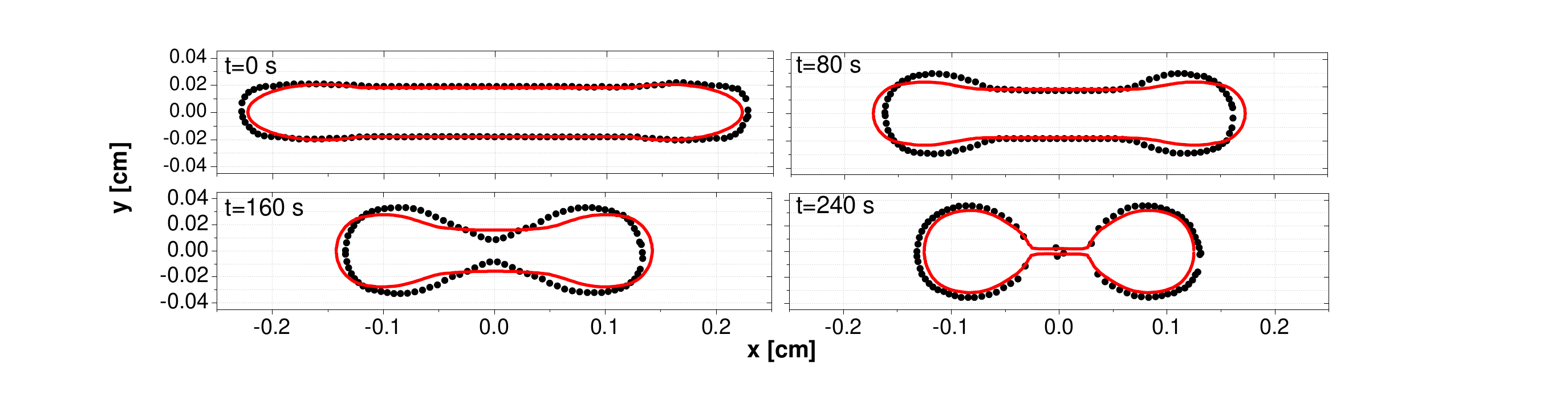}}
	\caption{{\it Numerical simulations versus experiments}: Time evolution of the contact line shape of two filaments leading to (a) one drop, and (b) two drops. The symbols correspond to experimental data and the solid lines to the simulations.}
	\label{fig:cont_num}
\end{figure}

In order to validate the numerical scheme, we simulate the evolution of actual filaments and compare the numerical contact line profiles at different times with those from the experiments. This comparison can be seen in Fig.~\ref{fig:cont_num} for two different filaments, one which ends up into a single drop and another one that breaks up into two drops. Although these filaments have almost the same width and they differ on their initial length on about $20\%$, the respective aspect ratios, $\Delta=L_i/w$, are different enough to yield completely diverse results. Note that the end shapes of the initial filaments studied here are pretty peculiar, since they are the result of the breakup process that occurs at the intersections of two very long filaments. For simplicity, we emulate these shapes in the simulations by ellipsoidal caps whose lengths and widths ($ \gtrsim w$) correspond to the measured experimental values.

\begin{figure}[htb]
        \includegraphics[width=0.9\linewidth]{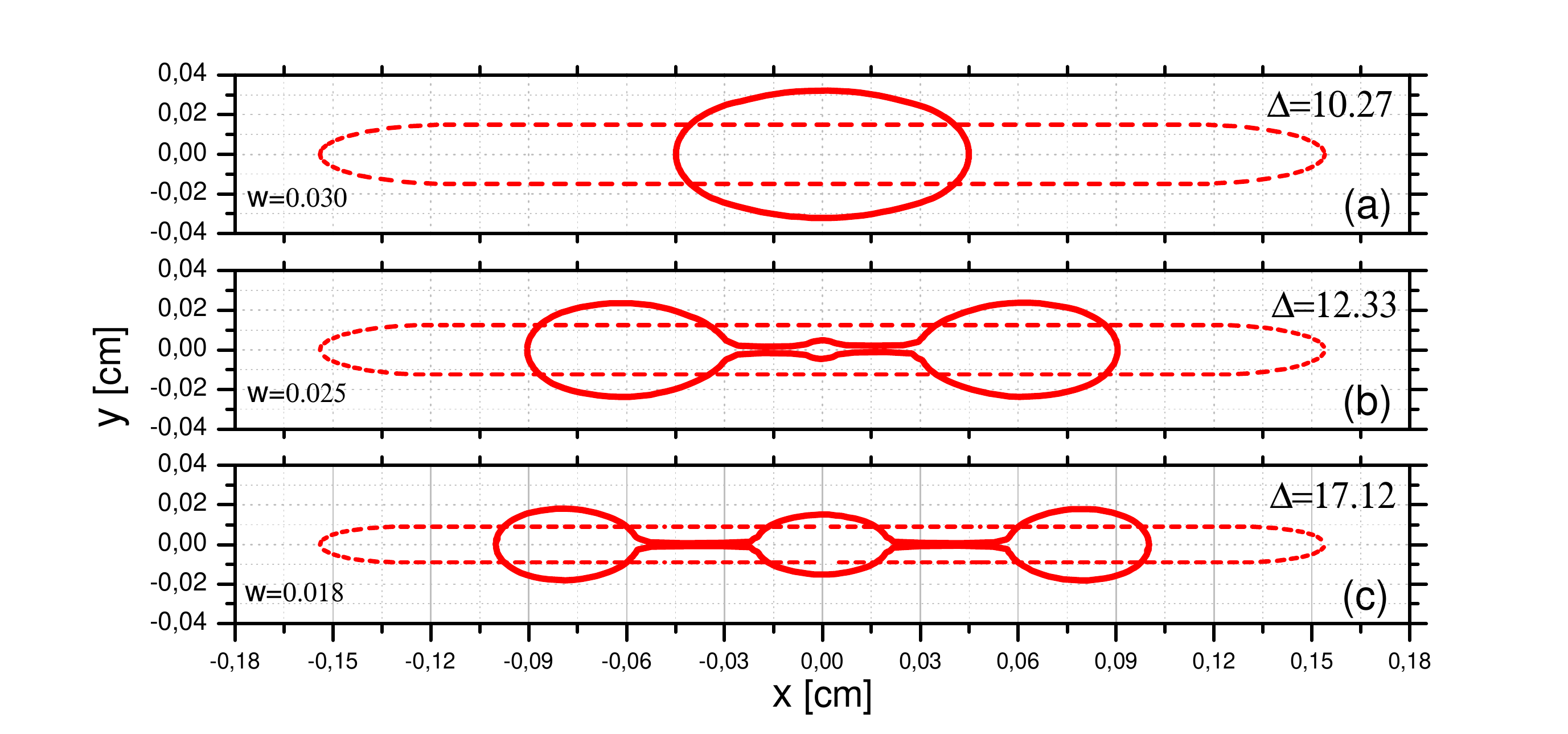}
	\caption{Initial (dashed lines) and final (solid lines) footprints of three filaments with the same initial length, $L_i=0.30825$~cm, and different widths: (a) $w=0.030$~cm ($\Delta=10.27$), b) $w=0.025$~cm ($\Delta=12.33$), and (c) $w=0.018$~cm ($\Delta=17.12$). The resulting number of drops is in agreement with the model predictions (see e.g. Fig.~\ref{fig:ndrops}).}
	\label{fig:num_exp}
\end{figure}

Considering the fact that it is not easy to generate filaments of prescribed width with a precision less than $0.05$~cm and that the results are strongly dependent on the aspect ratio $\Delta=L_i/w$, the numerical simulations may be used as an important tool to contrast the predictions of the model developed in Section~\ref{sec:model}. Thus, we perform a series of runs with the same initial length, $L_i$, and vary the width, $w$.  We take a typical $L_i$ equal to the average of a group of experiments, $L_i=0.30825$~cm, and vary $w$ in the range $(0.017,0.035)$~cm. Then, inside the corresponding $\Delta$-range, namely $(8.8,18.1)$, the model predicts the transition from one to two drops at  $\Delta_2=11.08$, and from two to three at $\Delta_3=16.62$, respectively. The results shown Fig.~\ref{fig:ndrops} by solid circles for six values of $\Delta$ are in agreement with this prediction. In particular, in Fig.~\ref{fig:num_exp} we show the evolution of three filaments with the same initial length and different withs, where the diverse aspect ratios lead to different number of drops, as predicted by the model of the previous sections. While the simulations are not able to end in a breakup of the connecting bridge, which remains forever, the results of the simulations confirm that the model of the previous sections has the essential features needed to explain the the final pattern of drops. 

\section{Summary and conclusions}
\label{sec:conclu}
We study the complete breakup processes of a set of two sets of parallel silicone oil (PDMS) liquid filaments crossing each other perpendicularly. They are deposited on a glass substrate previously coated with a fluorinated solution to achieve partial wetting conditions. Interestingly, the resulting rectangular grid yields two  different types of drops, depending on whether they are formed at the filament intersections or along the filaments themselves. We find that they differ on the shape of the footprint as well as on the contact angle distribution along the periphery, $\theta (\phi)$, where $\phi$ is polar angle. The drops at the intersections have rhomboidal--like shape (see Fig.~\ref{fig:drop_int}a), while those along the filaments have an ellipsoidal--like shape (see Fig.~\ref{fig:drop_fil}a). These differences have a natural implication on the contact angle distribution. In the first case, $\theta$ is maximum at $\phi=k\pi/4$ ($k=1,3,5,7$), and minimum at $\phi=k \pi/2$ ($k=0,1,2,3$) (see Fig.~\ref{fig:drop_int}b), while in the second case the maximum is at $\phi=k \pi/2$ ($k=1,3$) and the minimum at $\phi=k \pi$ ($k=0,1$) (see Fig.~\ref{fig:drop_fil}b). These extreme angles correspond to the advancing, $\theta_a$, and receding, $\theta_r$, contact angles of the hysteresis cycle~(\cite{rava_pof16}), respectively. Here, we develop an analytical solution for both drops in polar coordinates by solving the equilibrium equation corresponding to the balance of pressures inside the drop. This solution is expressed as series expansion in modified Bessel functions~\cite{rava_pof16}.

While the drops at a vertex of the rectangle are a consequence of four quasi simultaneous breakups, those formed at the sides are similar to those observed previously for a single filament. One advantage of our system is that one can compare at once four similar short filaments and see whether they all have the same behavior. Although, probably due to unavoidable initial perturbations, there is some variability in the number of drops formed in filaments with equivalent aspect ratios, it is restricted to precise bounds that can be predicted with our model.  

In order to find them, we developed a hydrodynamical model that accounts for all the possible types of breakups that can occur in a filament of given length and width, i.e. on its aspect ratio $\Delta=L_i/w$. We find that there are different ranges of $\Delta$ in which a certain number of drops are possible. The model considers the distance traveled by the filament end before a first neck shows up, and then evaluates the admissible values of the distance between the head and the breakup point consistent with a Stokes flow between the neck region and the head. The  model predictions are successfully compared with the experimental data (see Fig.~\ref{fig:ndrops}), and then it constitutes a useful tool to help designing grids with a desirable number of drops between intersections.

We can compare the value of the critical ratio $\Delta_2$ as given by our model with experimental data within the nanoscale. For instance, in~\cite{hartnett_lang15} the authors study the PLiD of a flat Ni strip on a $SiO_2$ substrate, which is melted by nanosecond pulses of laser beams. In Fig. 4a of that paper, they report the critical filament length for the transition from one to two drops as given by the best fit line of the experiments in the form $\ell_c=31.27 A^{1/2}$, where $A$ is the cross section of the cylindrical cap shape of the filament. This cross sectional area is assumed to be the same as that of the original rectangular flat strip. The rationale for this is that after some fast dewetting the system evolves to form a cylindrical cap of the same length of the original strip and this cap is prone to the formation of drops.  Since $A=R^2 ( \theta_s - \sin \theta_s \cos \theta_s )$ with $R=w/(2\sin \theta_s)$ being the cylinder radius and $\theta_s$ the {\it single} static contact angle, we have
\begin{equation}
\ell_c = 31.27 \frac {\sqrt{\theta_s - \sin \theta_s \cos \theta_s}}{2 \sin \theta_s} w = \Delta_{2}^{nano} w
\label{eq:hartnett}
\end{equation}
In~\cite{hartnett_lang15} they report $\theta_s=(69 \pm 8)^\circ$, so that $ \Delta_{2}^{nano}=15.9\pm 3.06$. On the other hand, the value predicted by our model is $\Delta_{2}=9.40$, where we used $\gamma_{Ni}=1.78$~N/m, $\mu_{Ni}=4.61$~mPas, and $\theta_a=\theta_r=\theta_s$. This is not so bad an estimation considering several facts. First, besides the experimental error in the determination of $\theta_s$ (which is certainly a difficult task), we must also bear in mind that $\ell_c$ in ~\cite{hartnett_lang15} refers to the length of the flat strip that evolves into a cylindrical cap when melted by the laser heating, and not to the length of the cylindrical filament itself as we mean here. It is known that the ends of the strip retract when melting and evolving into the cylindrical filament, so that one must expect that $\ell_c$ is actually larger than the filament length $L_{2}$, which justifies a larger value of the proportionality constant. If one estimates this retraction of the order of $L_d$, which seems reasonable, the results fit fairly well. Moreover, a second factor to be taken into account is a consequence of the previous one. If the filament retracts, the mass at the ends is relocated in a filament of shorter length. When this happens it is not exactly true that the cross sectional area of the original strip is equal to that of a cylindrical cap of equal length, since the the cylindrical cap is shorter and the central region has increased its mass per unit length. Then, the real value of $w$ can be expected to be larger than the value given by the assumption used in~\cite{hartnett_lang15}, which further reduces the value of the aspect ratio of the initial filament as our theory is pointing out.  

Finally, we also numerically simulate the time evolution of the filament by solving the complete Navier--Stokes equation, including both a slip condition and a contact angle dependence on contact line velocity. These boundary conditions are certainly constitutive relations for our physical system, so we employ the laws derived for it in a previous work~\cite{rava_pre17}. The time evolution is firstly validated by comparison with experimental results (see Fig.~\ref{fig:cont_num}). Even though the numerical scheme is not able to completely describe the breakup process, the simulations are still useful to give the trend to the formation of a given number of drops. The results as shown in Fig.~\ref{fig:num_exp} are successfully compared with both the experiments and the model (see Fig.~\ref{fig:ndrops}). Therefore, the simulations with the corresponding conditions at the contact line are useful to describe some details of the whole dynamics in addition to the final drop pattern.

In summary, we have shown that experiments can be dealt satisfactory with full simulations and a simple physical model of the rupture process. Moreover, the model predicts the time evolution as the succession of breakups seen in experiments, a factor not taken into account in previous infinite length models which assume simultaneous breakups. It also leads to predictions that can be related to previous attempted approaches and yields an useful tool to estimate the number of drops resulting from a given filament by considering the wetting properties of the liquid/substrate interaction in the submillimetric scale while, within the experimental uncertainties, it can be used as a first approach in nanometric scales.

\acknowledgments
The authors acknowledge support from Consejo Nacional de Investigaciones Cient\'{\i}ficas y T\'ecnicas (CONICET, Argentina) with grant PIP 844/2012 and Agencia Nacional de Promoci\'on Cient\'{\i}fica y Tecnol\'ogica (ANPCyT, Argentina) with grant PICT 931/2012.


\end{document}